\documentclass[review]{elsarticle}

\usepackage{color}
\setcitestyle{numbers,square}
\usepackage{epsfig}
\usepackage{color}
\usepackage{times}
\usepackage{mathtools}
\usepackage{amsmath}
\usepackage{amssymb}
\usepackage{indentfirst}
\usepackage{graphicx}
\usepackage{bm}
\usepackage{longtable}
\usepackage{wrapfig}
\usepackage{comment}
\usepackage{geometry}
\newcommand{\fg}[1]{\mbox{\pmb{$#1$}}}
\newcommand{\bey}{\begin{eqnarray}}
\newcommand{\eey}{\end{eqnarray}}

\newcommand{\sg}{\sigma}
\newcommand{\fsg}{\fg \sigma}

\newcommand{\bec}{\begin{center}}
\newcommand{\eec}{\end{center}}

\usepackage{numcompress}

\begin{document}

\setcounter{tocdepth}{2}
\baselineskip   15pt 
\belowdisplayskip14pt 
\belowdisplayshortskip10pt 
\renewcommand{\thefootnote}{\fnsymbol{footnote}}

\begin{frontmatter}

\title{Virtual melting and cyclic transformations between amorphous Si, Si I, and Si IV in a shear band}



\author[mymainaddress]{Hao Chen\corref{mycorrespondingauthor1}}
 \cortext[mycorrespondingauthor1]{Corresponding author}
 \ead{haochen_mech@ujs.edu.cn}

\author[mysecondaddress,mythirdaddress]{Valery I. Levitas\corref{mycorrespondingauthor2}}
 \cortext[mycorrespondingauthor2]{Corresponding author}
 \ead{vlevitas@iastate.edu}

\address[mymainaddress]{School of Mechanical Engineering, Jiangsu University, Zhenjiang 212013, P.R. China}

\address[mysecondaddress]{Iowa State University, Departments of Aerospace Engineering and Mechanical Engineering, Ames, Iowa 50011, USA}

\address[mythirdaddress]{Ames National Laboratory, Division of Materials Science and Engineering,  Ames, IA, USA}

\begin{abstract}
Virtual melting (VM) as alternative deformation and stress relaxation mechanisms under extreme load
is directly validated by molecular dynamics (MD) simulations of the simple shear of single crystal Si I at a temperature 1,383 K below the melting temperature. The shear band consisting of liquid Si is formed immediately after the shear instability while stress drops to zero. A thermodynamic criterion for VM, which depends on the ratio of the sample to shear band widths, is derived analytically and confirmed by MD simulations. With further shear, the VM immediately transforms to a mixture of low-density amorphous a-Si, Si I, and IV, which undergo cyclic transformations a-Si$\leftrightarrow$Si I,  a-Si$\leftrightarrow$Si IV, and Si I$\leftrightarrow$Si IV with volume fraction of phases mostly between 0.2 and 0.4 and non-repeatable nanostructure evolution. Such cyclic transformations produce additional important carriers for plastic deformation through transformation strain and transformation-induced plasticity due to volume change, which may occur in shear bands in various material systems but missed in experiments and simulations.

\end{abstract}


\end{frontmatter}

\section*{Introduction}
The phenomenon of the virtual melting (VM) much below the melting temperature $T_m$ as the mechanism of crystal-crystal and crystal-amorphous phase transformations (PTs) via intermediate melt was suggested in \cite{levitas2004solid, levitas2005crystal}. It was justified using simple thermodynamic estimates under zero \cite{levitas2004solid} and constant \cite{levitas2005crystal} pressure and applied to organic crystals, geological materials, Si, etc. The main thermodynamic driving force at zero pressure was the complete relaxation of the internal stresses caused by the transformation strain during crystal-crystal PT when other stress relaxation mechanisms (such as generation and motion of dislocations and twinning) are suppressed.
Since melt is unstable at stress-free conditions, it solidifies during a short time; that is why it is called virtual melting.
In \cite{levitas2005crystal} the VM was caused by crossing metastable continuation of the melting line to low temperature for materials with reducing melting temperature with growing pressure, like in ice\cite{mishima1984melting, mishima1996relationship}, quartz\cite{hemley1988pressure},  Si and Ge\cite{brazhkin1996lattice, brazhkin1997nature}. VM has been found at a temperature 4,000 K below the melting temperature under a very high strain rate ($10^9-10^{12}/s$) uniaxial compression of Al and Cu single crystals
using molecular dynamic (MD) simulations \cite{2012Virtual}. After completing the melting and reaching the hydrostatic state,
crystallization occurs because the relaxation of nonhydrostatic stresses eliminates the thermodynamic driving force for melting.
A much lower reduction in melting temperature was observed in MD simulations for Cu single crystal in the shock wave in \cite{ravelo2006directional, an2008melting, budzevich2012evolution}.

A nonequilibrium but stable disordered phase was obtained in MD simulations near the friction surfaces under high normal stress
in Si and diamond \cite{2018Shear,deringer2021origins, pastewka2011anisotropic, reichenbach2021solid, hu2023amorphous, zhao2023amorphization}.  It differs from the VM or virtual amorphous phase because it does not disappear while sliding is continued.

There are numerous experiments on melting/amorphization in shear bands in Si under static uniaxial compression\cite{he2016situ}, scratching\cite{minowa1992stress, zhang2015novel}, and dynamic loadings\cite{ZHAO2016519, zhao2021amorphization, zhao2016directional}.
In \cite{he2016situ}, amorphization occurred due to the accumulation of defects, which also was the case in MD simulations \cite{chen2019amorphization}.
In \cite{ZHAO2016519, zhao2021amorphization}, VM melting was considered as the main amorphization mechanism; it was treated using a thermodynamic
approach \cite{levitas2004solid} extended by the concept of the modified transformation work \cite{zarkevich2018lattice}.

To summarize, while shear-induced amorphization of Si under high pressure was observed in experiments and MD simulations,
there is no proof that it occurs via VM. Reasons for shear-induced VM and amorphization, like elastic or phonon instability or thermodynamic melting, are unclear. Because melting in a shear band occurs under drastically reduced
shear stresses, a strict thermodynamic approach for shear-induced melting is lacking.
With the classical thermodynamic approaches \cite{grinfel1991thermodynamic},
reduction in the melting temperature of the stress-free solid caused
by deviatoric stresses was estimated to be 1 K only.
Here, we utilize MD and develop a thermodynamic approach to reveal the main atomistic features and criteria for the simple shear-induced VM under normal pressure. We also revealed and rationalized the transformation of the VM to the mixture of low-density amorphous a-Si,  Si I (cubic diamond), and Si IV (hexagonal diamond) with cyclic PTs between these three phases.

 Vectors and
tensors are denoted in boldface type;  $\fg m \fg n= \left(m_{i} \, n_{j}\right) $ is the dyadic products of the vectors $\fg m $ and $\fg n $; ${\fg A} \, {\fg \cdot} \,
{\fg B} \, = \, \left(A_{ij} \, B_{jk}\right) \,$ and $\, {\fg A} \, {\fg :} \,
{\fg B} \, = \, A_{ij} \, B_{ji}\,$ are the contraction of tensors
over one and two nearest indices. Superscripts $\; - 1 \,$ and $T$ denote
inverse operation and transposition, respectively. Averaged over the volume parameters are designated with the bar, i.e., for averaged
shear stress $\bar{\tau}$ and strain $\bar{\gamma}$.

\section*{Results}

{\bf MD simulations}

\begin{figure} [!htbp]
 \centering
 \includegraphics[width=\textwidth]{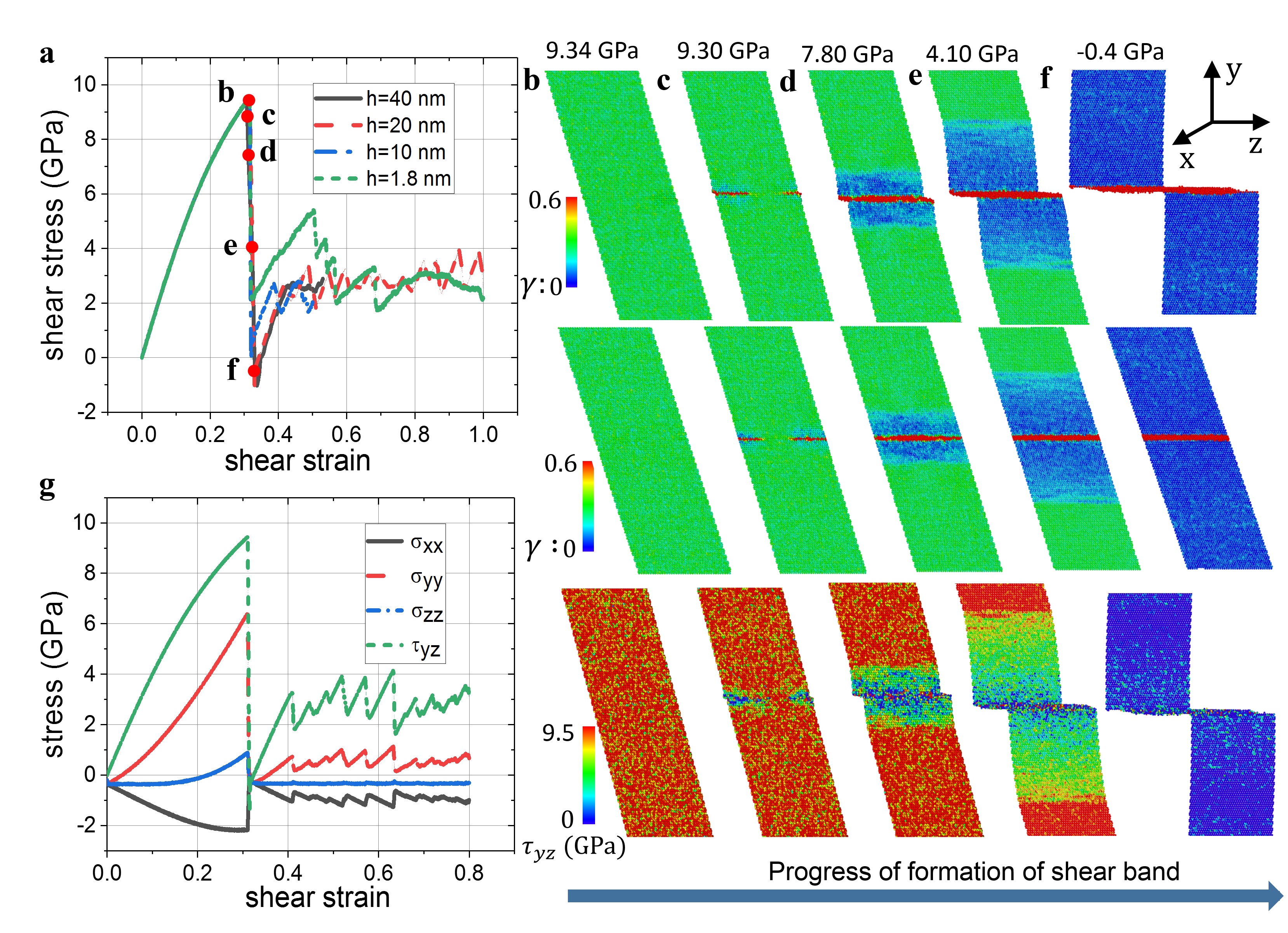}
    \caption{
    {\textbf{Stress and strain fields in Si during shear.} (a) Averaged shear stress-strain relations for different sample widths. The red dots indicate the selected loading state for presenting the shear strain and stress snapshots. (b-f) The shear strain $\gamma$ and stress $\tau_{yz}$ snapshots for corresponding states marked with the red dots in (a). The shear stain $\gamma$ is presented both in the actual configuration corresponding to the Lagrangian description of the initial rectangular box (top) and the wrapped configuration corresponding to the Eulerian box in the space coinciding with the actual configuration in (b) (middle). (g) Stress tensor components versus shear strain for the sample width of $20\, nm$. }
    }
    \label{fig:plotS1}
\end{figure}

\begin{figure} [!htbp]
 \centering
 \includegraphics[width=\textwidth]{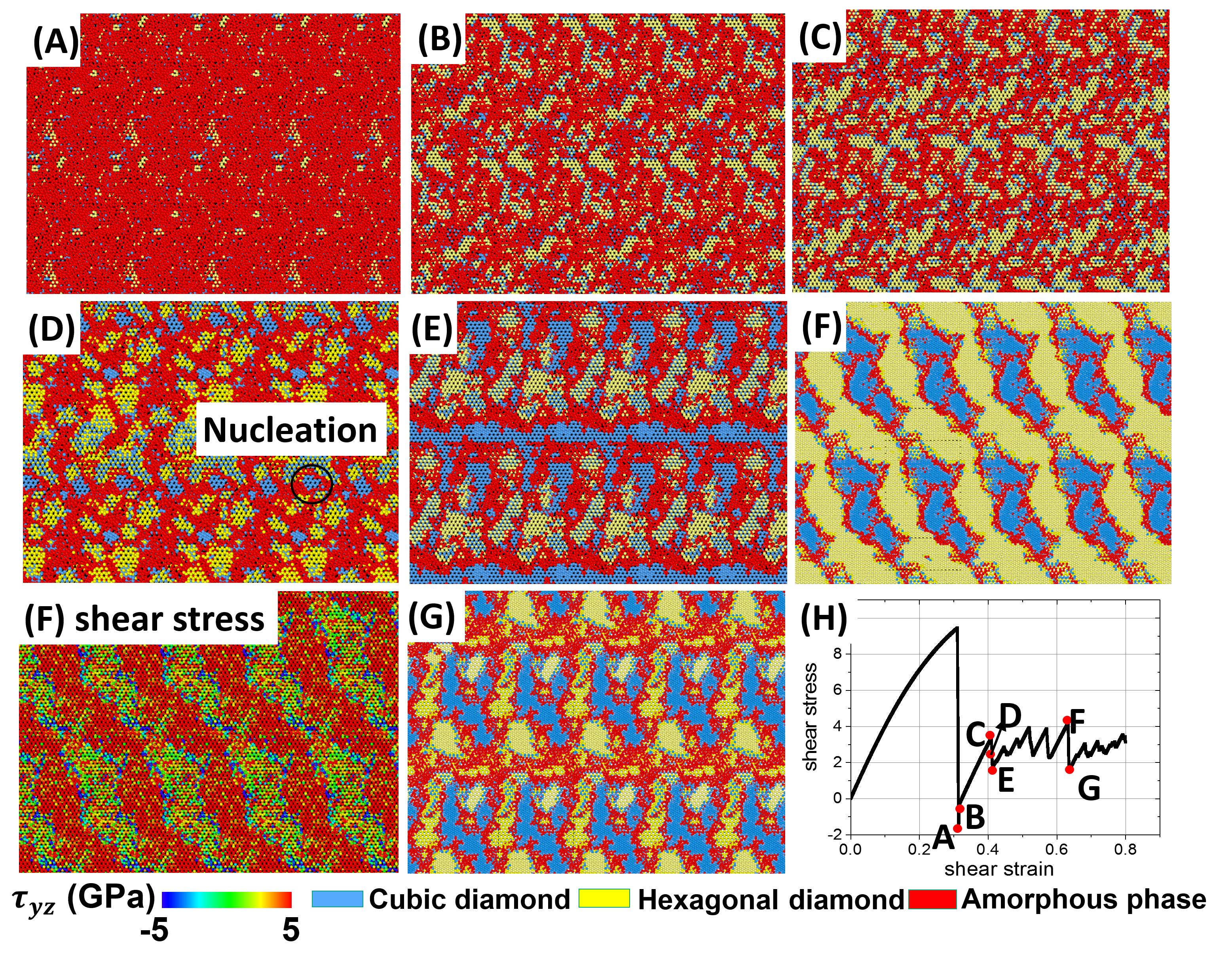}
    \caption{
    {\textbf{Cyclic a-Si$\leftrightarrow$Si I, a-Si$\leftrightarrow$Si IV, and
 Si I$\leftrightarrow$Si IV phase transformations in the shear band during the shear loading.}
 Snapshots of cubic diamond, hexagonal diamond, and amorphous phases are presented in  (A-G), corresponding to the points in the shear stress-strain curve in (H).
 For point F in (H), the stress distribution is also presented.
 }
    }
    \label{fig:plotS2}
\end{figure}

MD simulations of the behavior of
the perfect single Si I crystal under the shear displacement $u$  along $\langle 110 \rangle$ direction on $\{ 111 \}$ plane at constant width
$h$ and temperature $300 K$ were performed.  Effects of the free surfaces on the formation of the shear band were excluded by employing periodic boundary conditions along all three cubic directions. No external stresses were applied to avoid reduction of the melting temperature of Si $T_m=1683\, K$ due to pressure. The sample width $h$ has been varied from 1.8 to 160 $nm$.
The averaged true (Cauchy) stress-shear strain $\bar{\gamma}=v/h$ curves and the local shear strain and true stress fields in the sample are presented in Fig. \ref{fig:plotS1}. The averaged shear strain rate is $\dot{\bar{\gamma}}=v/h =10^9/s$ for $v=10\,m/s$ and $h=10\, nm$ before the shear band appears.
With increasing shear up to point $b$, deformation represents uniform, simple elastic shear for any sample width. While the largest stress is the shear stress   $\tau_{yz}$, normal stresses are also essential, with large tensile $\sg_{yy}$ (in the vertical direction) and compressive out-of-plane $\sg_{xx}$; stress $\sg_{zz}$ is small. For all sample widths, instability leading to a disordered band starts at the same peak shear stress and corresponding strain for any sample thickness. Since the shear modulus at the peak shear stress is not zero, this is not elastic instability. Through the calculation of the Hessian matrix of the strained sample just before the instability point\cite{plimpton1995fast}, the instability mode is found to be that two atoms inside one primitive cell of Si move towards each other, which corresponds to the phonon instability of Si I \cite{grimvall2012lattice}. The elastic stress-strain curve and the instability
stress and strain are independent of the sample size (Fig. \ref{fig:plotS2}a).
The entire process of the shear band formation occurs at constant averaged shear strain (i.e., very fast) with a drastic drop of all stresses down to zero. This is a clear signature of the VM because amorphous solid supports shear and normal deviatoric stresses (see below).
Two atomic-thin disordered bands appear at different ends of the sample and propagate to each other until they coalesce into a single band.
At this stage, shear elastic strain and stresses drop almost to zero in the band and this drop propagates as a wave toward the upper and lower end of the sample. When the wave reaches the ends of the sample, shear stresses and elastic strain in the entire sample are practically absent.
Some overshot to small negative averaged shear stress is caused by dynamic process and viscosity of the melt.

\vspace{1.8 cm}
{\bf  Size-dependent thermodynamic criterion for shear-induced virtual melting}

Based on these results, a large-strain thermodynamic framework and criterion for shear-induced melting in a shear band is developed.
The only work where melting under non-hydrostatic loading was considered a thermodynamic process rather than the equilibrium across a solid-melt interface is \cite{2012Virtual}. However, melting in \cite{2012Virtual} occurred continuously and uniformly during uniaxial compression within some strain increment with the same axial stress at the beginning and completion of the melting. Since shear-induced melting occurs heterogeneously at  $\bar{\gamma}= const$ and stresses drop to zero, a different thermodynamic formulation is required.

\begin{figure} [!htbp]
 \centering
 \includegraphics[width=0.5\textwidth]{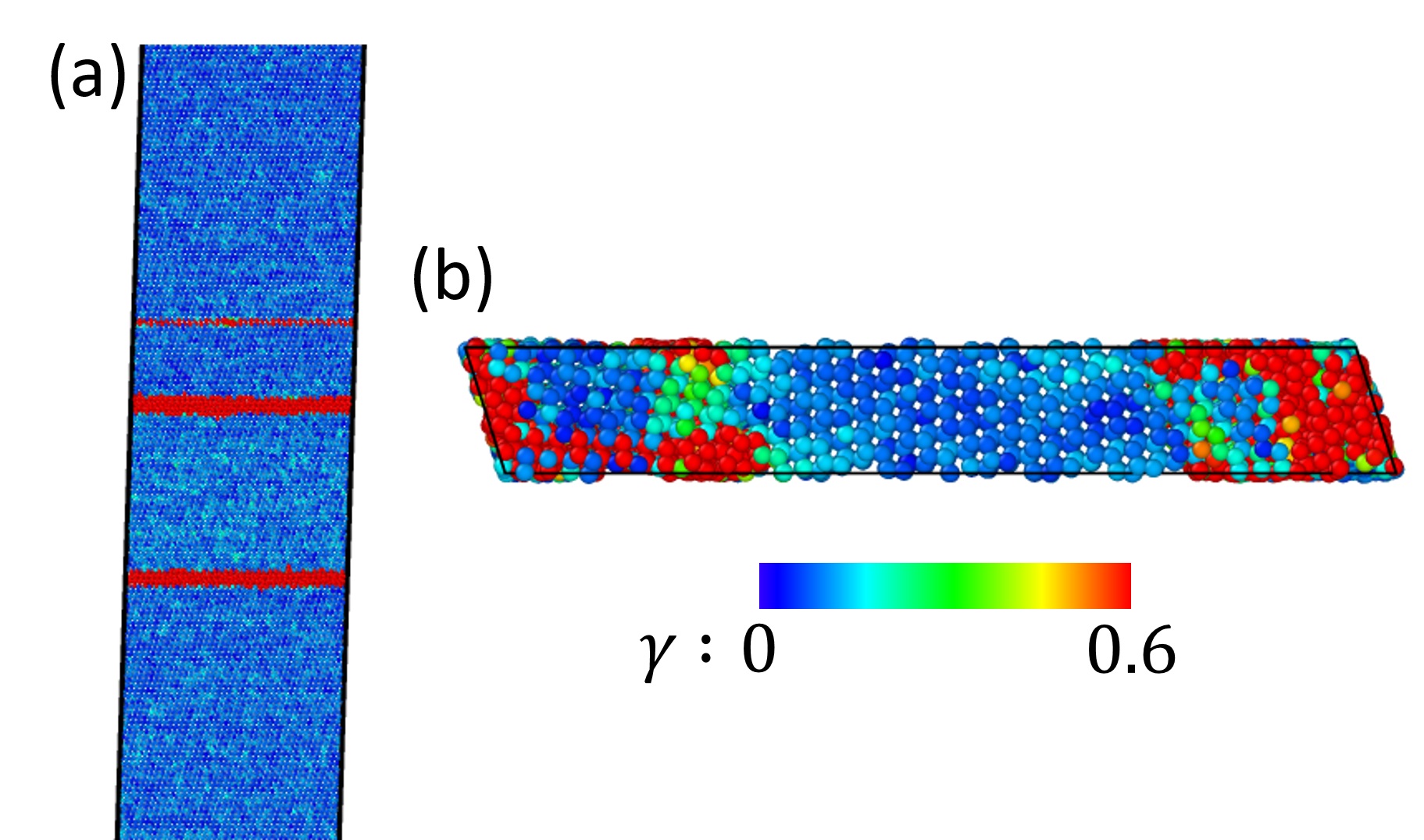}
    \caption{
    {\textbf{Effect of the sample width on shear banding.} (a) Multiple shear bands formed for a sample width of $80 nm$ under simple shear; (b) For a sample width of $1.8 nm$ (i.e., smaller than the critical width of $2.1 nm$), no shear band formation is found; discrete amorphous (red) regions are observed. 
 }
 }
    \label{fig:plotS4}
\end{figure}

The thermodynamic criterion for a PT in a volume $V_0$ of the entire sample bounded by the surface $S_0$ in the reference configuration is
determined by utilizing methods that we developed for the PTs in solids\cite{levitas2000structural, levitas2000structural2, levitas1998thermomechanical, levitas2021phase}
\bey
  \bar{X} \, := \int_{S_{0}}  \int^{\fg u_2}_{\fg u_1}  \fg p_0 \cdot d \fg u dS_{0}
 -   \int_{V_{0}} \, \left[ \left( \psi_2^e - \psi_1^e \right) + \, \Delta \psi^\theta (T) \right]d V_0
- \int_{\Sigma_0}  \Gamma_0
d \, \Sigma_0 \geq 0.
\label{ijss1-180}
\eey
Here $  \bar{X}$ is the total dissipation increment during the PT,  $\, \psi^e \,$ is the elastic energy per unit undeformed volume,  $\Delta \psi^\theta (T)$ is the change in the thermal part of the Helmholtz free energy per unit undeformed volume as a function of temperature $T$,
$ \Gamma_0$ is the interfacial energy per unit undeformed area, $\Sigma_0$  is the interface,
the indices 1 and
2 indicates the values before and after the PT.
Eq. (\ref{ijss1-180})  combines the second law of thermodynamics and the assumption that the entire dissipation occurs due to melting only, neglecting plastic dissipation in the disordered phase during the PT process.
Indeed, there is no dislocational or twinning plasticity within the crystal, and plastic flow occurs after sufficient disordering is developed, i.e., at the later stage and after the PT. We neglect this plasticity.
The PT starts at the peak of the true stress and ends at the lowest stress; if the lowest stress is zero (or slightly negative), melting occurs; otherwise, PT represents amorphization.

The convenience of application of the criterion (\ref{ijss1-180}) to the entire sample rather than to any part that includes shear is because the PT occurs at the fixed displacements $\fg u$ at the surface and   
the first integral in Eq. (\ref{ijss1-180}) disappears. Then $  \bar{X}$ depends on the initial (at the peak stress) and final (at the lowest stress for the same $\gamma$) states only and is independent of the disordering process.
For melting, we assume that elastic stresses and energy disappear at each point, i.e., $ \psi_2^e=0$; this increases the driving force in comparison with amorphization. Also, the crystal-amorphous interface has energy on the order of magnitude of the incoherent grain boundary, i.e., $1 \, J/m^2$, while the solid-melt interface has energy $(0.01 - 0.1) \, J/m^2$.
 Thus, for comparable $\Delta \psi^\theta$, the VM process is much more thermodynamically favorable than the amorphization.

 For a slab with the width $h$ and width of the shear band $h_b$, utilizing the homogeneity of all parameters within crystal and disordered phases before and after completing disordering,  the dissipation increment per unit band surface can be obtained from Eq. (\ref{ijss1-180}):
\bey
  \bar{X}_\Sigma= \bar{X}/\Sigma \, :=
 -    \,  \left( \psi_2^e - \psi_1^e \right) h - \, \Delta \psi^\theta h_b
- 2  \Gamma_0
 \geq 0.
\label{1-180-a}
\eey
Let us evaluate $\psi_1^e $ at the peak stress.
The first Piola-Kirchhoff stress $\fg P$ is related to the Cauchy stress  $\fsg$ by equation $\fg P=J^{-1} \fsg \cdot \fg F^{T-1}$,
where $\fg F$ is the deformation gradient and $J=det \fg F =dV/dV_0$ is the Jacobian determinant describing the ratio of elemental volumes in the actual and undeformed configurations.
For simple shear  with the magnitude $\gamma$ in direction $\fg m$ along the plane with the normal $\fg n$, $\fg F= \fg I + \gamma \fg m \fg n$, $J=det \fg F=1$,  and $\fg F^{-1}= \fg I - \gamma \fg m \fg n$. Then
 $\fg P= \fsg \cdot \fg F^{T-1}= \fsg \cdot (\fg I - \gamma \fg n \fg m)$. In particular, for a shear component of $\fg P$, $P_{mn}:=\fg m \cdot \fg P \cdot \fg n= \fg m \cdot \fsg  \cdot \fg n= \sg_{mn} = \tau$ (we took into account orthogonality of $\fg m$ and $\fg n$), i.e., the first Piola-Kirchhoff and the Cauchy shear stresses in the direction of shear $\fg m$  along the plane with normal $\fg n$ are the same. The elemental stress work per unit undeformed volume
$\fg P : d \fg F^{T}=  \fsg \cdot (\fg I - \gamma \fg n \fg m) : d \gamma \fg n \fg m=   \fsg  : d \gamma \fg n \fg m =  \sg_{mn} d \gamma = \tau  d \gamma $.
Since for elastic deformation before the instability strain $\gamma_c$, $\fg P : d \fg F^{T} = d \psi= \tau d \gamma $, then
$\psi_1^e=  \int^{\gamma_c}_{0} \tau d \gamma$, i.e., it is equal to the area below  $\tau -\gamma$
curve in Fig. \ref{fig:plotS1} (a).
Thus, despite the presence of normal stresses (see Fig. \ref{fig:plotS1}g), they do not produce mechanical work due to simple shear deformation and do not contribute to the thermodynamic driving force for shear-induced melting.

Dividing Eq. (\ref{1-180-a}) by $h$ and setting for melting  $\psi_2^e=0$, we obtain for the dissipation increment per unit undeformed volume
\bey
X:=  \bar{X}_\Sigma/h
=  \int^{\gamma_c}_{0} \tau d \gamma  - \frac{\Delta \psi^\theta  h_b+ 2  \Gamma_0}{h}\,
 \geq 0 . \qquad  
\label{1-180-b}
\eey
Eq.~\eqref{1-180-b} represents a size-dependent thermodynamic criterion for the VM in a shear band. Note that this is not a condition for the appearance of a critical nucleus because disordering occurs due to loss of stability and is, therefore, barrierless. This is a condition that disordering, which appeared due to material instability, could be transformed
into the liquid within the shear band. For the melting below thermodynamic melting temperature $\Delta \psi^\theta>0$,
i.e., elastic energy release should exceed the size-dependent threshold due to changes in thermal energy and surface energy.
 While it looks like for a large enough sample width $h$ this criterion can be satisfied for any material, this is not straightforward. For large $h$, multiple bands
 may appear (Fig. \ref{fig:plotS4}a), i.e., $h$ is approximately the distance between shear bands. Also, the VM may be completed before the stress relief wave from the shear band reaches the top and bottom of the sample, namely when the fully unloaded part covers width $h_1<h$.
 Then, the same thermodynamic procedure results in  Eq.~\eqref{1-180-b} with $h_1$ instead of $h$.

Change in the thermal energy can be approximated as $\Delta \psi^\theta (T)=\Delta \psi^\theta (300K)(1-(T-300)/(T-T_m))$ with
$\Delta \psi^\theta (300K)=41 \, kJ/mol= 3.39\, \times 10^9 \, J/m^3$ \cite{ZHAO2016519} (we took into account the molar volume of Si of $1.21 \times 10^{-5} m^3/mol$) and $T_m=1683\, K$.
In our simulations  $\psi_1^e=1.472 \times 10^9 \, J/m^3$,
$h_b= 0.9 \, nm$, and $\Delta \psi^\theta h_b= 3.051 \, J/m^2$. Assuming  $ \Gamma_0=0.1 \, J/m^2$,
 Eq.~\eqref{1-180-b} at 300 K takes a form
\bey
 h \geq 2.141 \, nm = 2.39 h_b.
\label{1-180-n}
\eey
For 0 K,  $\Delta \psi^\theta=  4.13\, \times 10^9 \, J/m^3$ and keeping the same other parameters, we obtain $  h \geq 2.59 \, nm = 2.88 h_b$.

To check this prediction, simulations with samples of several small $h$ are carried out. When sample width is reduced to $h = 1.8 \, nm$, shear stress dropped to 2 GPa only (Fig.\ref{fig:plotS1} (a)), and no shear band was formed (Fig. \ref{fig:plotS4} (b)). This is consistent with the prediction in Eq. (\ref{1-180-n}).
For the chosen parameters, the effect of the surface energy is negligible, and  Eq.~\eqref{1-180-b}
reduces to
\bey
\frac{h}{h_b} \geq
\Delta \psi^\theta /\int^{\gamma_c}_{0} \tau d \gamma .
\label{1-280}
\eey

{\bf Cyclic a-Si$\leftrightarrow$Si I, a-Si$\leftrightarrow$Si IV, and
 Si I$\leftrightarrow$Si IV phase transformations}

\begin{figure} [!htbp]
 \centering
 \includegraphics[width=\textwidth]{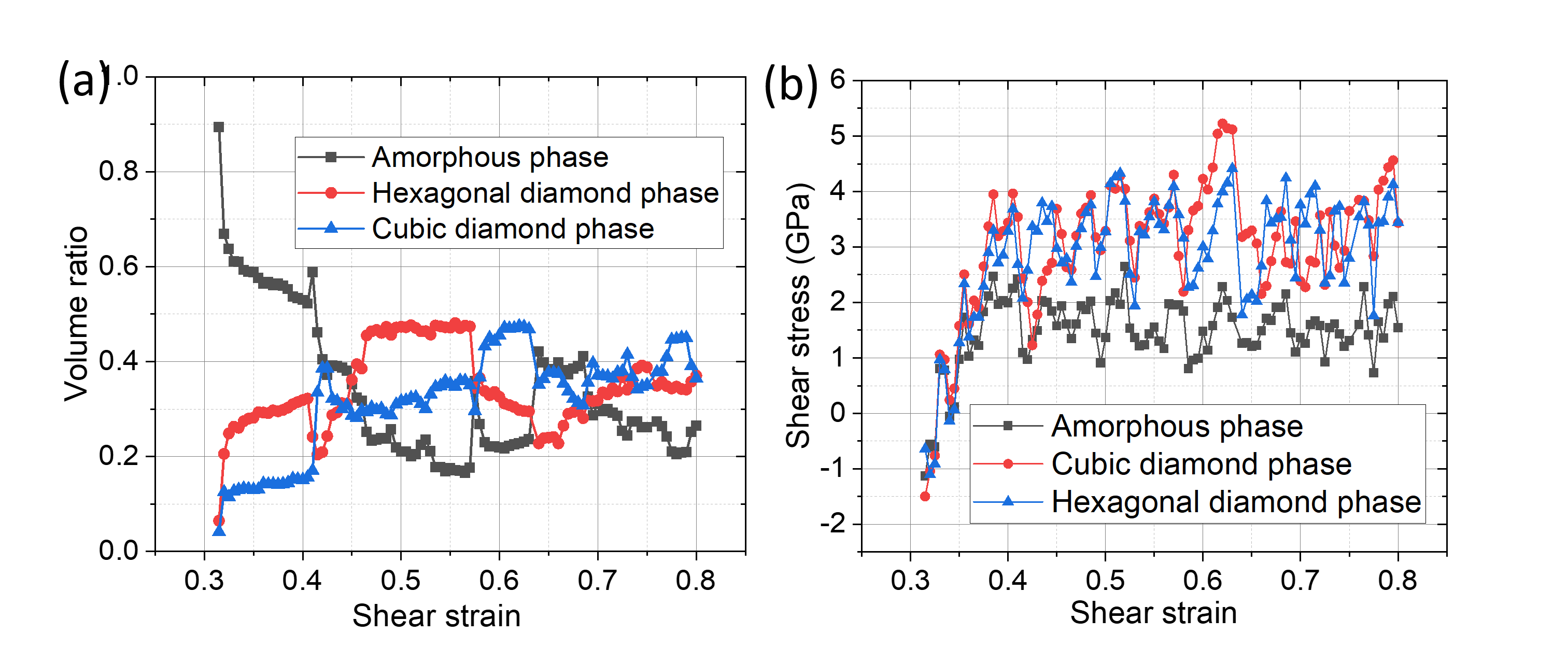}
    \caption{
    {\textbf{Kinetics of the cyclic shear-induced a-Si$\leftrightarrow$Si I, a-Si$\leftrightarrow$Si IV, and
 Si I$\leftrightarrow$Si IV phase transformations in a shear band and averaged shear stresses in phases.} (a) Evolution of volume fractions of each phase inside the shear band with increasing shear strain; (b) Evolution of average shear stress over each phase inside the shear band with increasing shear strain.
 }
 }
    \label{fig:plotS3}
\end{figure}

 After the formation of the shear band, the entire inelastic shear deformation is concentrated within the shear band, while the rest of the sample
 is in an elastic state.
Figs. \ref{fig:plotS2} and \ref{fig:plotS3} present evolution of the nanostructure (Fig. \ref{fig:plotS2}a-g), volume fraction of phases (Fig. \ref{fig:plotS3}a), shear stress distribution at the selected time instant (Fig. \ref{fig:plotS2}f), and shear stress-strain curves for each phase (Fig. \ref{fig:plotS3}b) and (with larger strain steps) mixture (Fig.  \ref{fig:plotS2}h).
 Several PTs occur within the shear band. Since melt is unstable at zero (or small) shear stresses, it transforms fast into an amorphous solid capable of supporting the shear stress. Based on a specific volume of a-Si and angular distribution function, this low-density a-Si is identical to that obtained by quenching at zero pressure from the melt. Simultaneously, Si IV and (in smaller amounts) Si I nanocrystals nucleate stochastically and, in most cases, homogeneously within the entire volume of the amorphous band instead of their heterogeneous nucleation at the boundary between a-Si and the initial Si I crystal or continuous growth of Si I crystal. This is one more essential difference between traditional and shear-driven PTs. Si I mostly appears at the boundary between Si IV and a-Si, but there are also Si I clusters within Si IV and a-Si. While the change in volume fraction of phases is monotonous for $0.31 <\gamma< 0.41$, stress oscillates and drops in the first cycle from 1 to 0 GPa, i.e., amorphous phase fluctuationally transforms for a very short time to VM. Stresses are the same in each phase for  $\gamma< 0.35$, implying that crystals are within the disordered phase, which is a transitional state between VM and the amorphous phase. For $0.41< \gamma< 0.42$, i.e., between points C and D, the next small increase and then large decrease in volume fraction of a-Si occurs along with an increase in volume fraction of Si I by 0.22 and decrease in volume fraction of Si IV by 0.11. This process is accompanied by a drastic change in the morphology of phases, coarsening of Si I and Si IV crystals, and finally, formation at point E of a layered structure
 including two continuous a-Si and Si I bands. Stress in a-Si drops to 1 GPa, which along with the formation of continuous a-Si bands leads to stress drop in crystalline phases and mixture. Between points C and E, all possible PTs, a-Si$\leftrightarrow$Si I, a-Si$\leftrightarrow$Si IV, and
 Si I$\leftrightarrow$Si IV, occur in both directions in different regions, including nucleation and continuous growth.
   Such PTs include Si I$\leftrightarrow$Si IV PT via intermediate a-Si.
  PT from stable at normal conditions Si I to the metastable Si IV during plastic deformation experiment is reported in the literature\cite{he2016situ, tan1981diamond, kulnitskiy2016mutual, blank1996crystallogeometry};
Si I and IV have the same atomic volume, and the transformation strain between these lattices consists of shears, similar to plastic shears. In particular, one of the Si I$\rightarrow$Si IV PT mechanisms is related to the motion of partial dislocations producing different stacking sequences, which can be considered an additional plastic relaxation mechanism in the current work.
 In experiment \cite{he2016situ} Si I within the shear band transforms to Si IV and, after accumulation of significant dislocation density, to a-Si.   Note that the width of the shear band here is an order of magnitude smaller, and the strain rate is many orders of magnitude larger than in\cite{he2016situ}, which explains the difference in the processes.
 The volumetric strain of 7.5\% \cite{deb2001pressure} during amorphization/crystalization of Si I and IV within the shear band creates internal normal stresses. Based on the solution in \cite{levitas2022resolving}, they produce large transformation-induced plasticity (TRIP), which was used to quantitatively explain large PT-induced shears during deep-focus earthquakes. Note that TRIP strain is independent of the sign of the volumetric strain, i.e., it occurs in the direction of shear stresses for both direct and reverse PTs.
TRIP explains a paradoxical result: despite the PT from weaker a-Si to stronger crystalline phases, shear stress relaxes.
At the atomistic level, TRIP cannot be distinguished from traditional plasticity because both are caused by local shear stresses.

With further straining, cyclic PTs between three Si phases occur without any specific repeating morphological patterns and
patterns in volume fraction evolutions. Oscillations in stresses in phases are much more frequent than in volume fractions of phases, caused
by the elemental act of plastic deformation in amorphous materials related to a series of irreversible rearrangements in small shear transformation zones\cite{argon1982mechanisms, delogu2008identification}. These rearrangements are similar to thermal fluctuations, which give numerous chances for PT
to the stable or metastable crystalline phases to occur under prescribed shear. Other reasons for oscillations and non-repeatable
patterns are caused
 by changes in crystal orientations and morphology of phases. Shear stress in a-Si cycles mostly between $\sim$1 and 2 GPa, and in crystalline Si, mostly between $\sim$2 and 4 GPa, with some outliers. The volume fractions of phases cycle mostly between 0.2 and 0.4, with some outliers.
 At point F, Si IV dominates over Si I
with a small amount of a-Si at the interfaces between Si I and IV and grain boundaries. Shear stress in Si I exceeds 5 GPa. However, such a system is unstable because it cannot accommodate large plastic shear. That is why both Si I and IV partially transform to a-Si with a large stress drop caused by TRIP, producing two bands of mostly a-Si as a carrier for further plastic straining.

Theoretical kinetic equation for strain-induced PTs between three phases\cite{levitas2006kinetics} has a stationary solution for large plastic strains under constant stresses. Here, we have cyclic stochastic direct-reverse PTs between a-Si, Si I, and Si IV, which challenge the existing theory and will be addressed in the future.

\section*{Discussion}

In the letter, VM as an alternative deformation and stress relaxation mechanism to dislocation and twinning plasticity
is directly validated by MD simulations for the simple shear of single crystal Si I at a temperature 1,383 K below the melting temperature.
Since pressure was not applied, we did not utilize a reduction in melting temperature with pressure, in contrast to previous papers\cite{deb2001pressure, hemley1988pressure, brazhkin1997nature}.
Si I lattice loses its stability due to phonon instability, resulting in the formation of a few $nm$ thick disordered shear band.
Since stresses drop down to zero, the disordered phase is a VM. The advanced thermodynamic criterion for VM as a deformation-transformation process, which depends on the ratio of the sample to shear band widths, $h/h_b$, is derived analytically and confirmed by MD simulations.
The main driving force for the VM is the reduction of elastic energy in the entire sample down to zero, which is utilized to overcome
barrier due to the change in thermal and surface energy within the shear band; this is the reason for the dependence of the criterion on $h/h_b$.
With further shear, the VM  immediately transforms to a mixture of low-density amorphous a-Si, Si I, and IV, which undergo cyclic transformations a-Si$\leftrightarrow$Si I,  a-Si$\leftrightarrow$Si IV, and Si I$\leftrightarrow$Si IV with volume fraction of phases mostly between 0.2 and 0.4 and non-repeatable nanostructure evolution. Such PTs include Si I$\leftrightarrow$Si IV PT via intermediate a-Si.
 Surprisingly,
Si IV and Si I nanocrystals nucleate stochastically and, in most cases, homogeneously within the entire volume of the amorphous band instead of their heterogeneous nucleation at the boundary between a-Si and the initial Si I crystal or continuous growth of Si I crystal.
Such a crystallization is plastic strain-induced. The atomic rearrangements in the small shear zones of amorphous material, as the primary
mechanism of plastic deformation in them,  are similar to thermal fluctuations, which give numerous chances for PT
to the crystalline phases to occur under a prescribed shear. However, when the total volume fraction of crystalline phases exceeds 0.8, this suppresses plastic flow in a-Si, shear stress grows and causes jump-like transformation to a-Si and Si I$\leftrightarrow$Si IV PTs in different regions of the band. Si I$\leftrightarrow$Si IV PTs produce shear transformation strain (in some cases, through the propagation of the partial dislocations).
The volumetric strain of 7.5\%  during amorphization/crystallization of Si I and IV generates internal normal stresses, which combined with applied shear stress produce a large TRIP.
Both transformation strain and TRIP serve as additional mechanisms of stress relaxation and allow accommodating prescribed plastic 
shear.
 TRIP strain is independent of the sign of the volumetric strain, i.e., it occurs for both amorphization and crystallization.
TRIP explains a paradoxical result:  despite the PT from weaker a-Si to stronger crystalline phases, shear stress relaxes.
Cyclic stochastic direct-reverse PTs between a-Si, Si I, and Si IV challenge the existing theory for strain-induced PTs.

Similar phenomena under shear may occur in various strong crystals, like Ge, superhard cubic and hexagonal diamonds and BN,
SiC during friction and wear, surface treatment (polishing, turning, scratching, etc.), intense dynamic loading, and shock waves.
Due to the transient character of the VM and high frequency of cycling, these phenomena may be  missed in experiments and
also in simulations.

{\bf Acknowledgements:}
 HC acknowledges support from the National Natural Science Foundation of China (No.52005186). VIL acknowledges support from NSF (CMMI-1943710, DMR-2246991, and XSEDE MSS170015), and Iowa State University (Vance Coffman Faculty Chair
Professorship and Murray Harpole Chair in Engineering).

{\bf Author contributions:} HC  performed MD simulation. 
VIL  developed theoretical model.  HC and VIL  prepared the manuscript.

{\bf Competing interests:}
The authors declare no competing interests.

{\bf Data availability:}
The data supporting this study's findings are available from the corresponding authors upon request.

\bibliography{mybib}

\section*{Methods}

In this work, classical molecular dynamics simulations were performed using the LAMMPS package \cite{plimpton1995fast}. The employed interatomic force field for the interactions between Si atoms was from the Stillinger-Weber (SW) potential \cite{balamane1992comparative}. This potential has been demonstrated to be successful in describing not only both the crystal and amorphous phase of silicon but also their mutual transformation \cite{vink2001fitting,demkowicz2004high}. The latest machine learning potential was also employed \cite{bartok2018machine}, and the results are the same with the SW potential.
 Here, the virial stress is the averaged stress of the whole atomistic system and is thus considered  equivalent to the Cauchy stress\cite{subramaniyan2008continuum}.

\newpage

{\bf Supplementary Material}

The same as at the beginning of paper, because it should be separate document.

{\bf Supplementary discussion}

{\bf Alternative derivation of the thermodynamic criterion for shear-induced virtual melting}

Alternatively and to get a better confidence and feeling of the melting criterion, we can apply Eq. (\ref{ijss1-180}) to the shear band only, because dissipation occurs within the shear band only.
Then, neglecting the change in kinetic energy between the initial and the final state of the VM, and applying the Gauss theorem, we obtain
\bey
 \int_{S_{b}}  \int^{\fg u_2}_{\fg u_1}  \fg p_0 \cdot d \fg u dS_{b} =
  \int_{V_{b}} \int^{\fg F_2}_{\fg F_1} \fg P : d \fg F^{T} d V_b = \int_{V_{b}}\int^{  \gamma^b_2}_{  \gamma^b_1} \tau  d \gamma^b d V_b;
\label{2-180}
\eey
\bey
 \bar{X}=\int_{V_{b}}\int^{ \gamma^b_2}_{  \gamma^b_1} \tau  d \gamma^b d V_b  +  \int_{V_{b}} \, \left(  \psi_1^e  - \, \Delta \psi^\theta \right)d V_b
- \int_{\Sigma_0}  \Gamma_0
d \, \Sigma_0 \geq 0,
\label{2-181}
\eey
and
\bey
X=\int^{ \gamma^b_2}_{\gamma^b_1} \tau_b  d \gamma^b   +   \, \left[  \psi_1^e  - \, \Delta \psi^\theta \right]
-\frac{2 \Gamma_0}{h_b}\geq 0,
\label{2-182}
\eey
where $ \gamma^b $ and $\tau_b$ are the averaged shear strain and stress in a shear band. To evaluate the integral in Eq. (\ref{2-182}), one needs to know the constitutive equation $\tau_b  ( \gamma^b)$ during the entire transformation process from crystal to the VM, which
is unknown. For neglected change in the kinetic energy, the continuity of shear stresses at the shear band-crystal interface and in the entire crystal is valid, which follows from the mechanical equilibrium equation.
 Then $\tau_b=\tau$. During the transformation process that occurs at $\gamma_c=const$
\bey
\gamma_c=  c \gamma_b + (1-c) \gamma \quad \rightarrow \quad d \gamma_b= \frac{c-1}{c} d \gamma; \qquad  c:=h_b/h,
\label{2-183}
\eey
where $c$ is the volume or linear fraction of the disordered band in a sample, and $\gamma$ is the shear strain in the crystalline part of the sample, which is considered to be homogeneous due to the homogeneity of the shear stress. The incremental equation follows from the first Eq. (\ref{2-183}) since $\gamma_c$ and $c$ are constant during the transformation process. At the beginning of the transformation  $\gamma_b=  \gamma = \gamma_c$,  at the end $\gamma =0$ and $\gamma_b= \gamma_c/c$, which determines integration limits in
the work integral in Eq. (\ref{2-182}). Based on these assumptions, we can evaluate the work integral
\bey
\int^{ \gamma_c/c}_{\gamma_c} \tau_b  d \gamma^b  =\frac{c-1}{c} \int^{ 0}_{\gamma_c} \tau   d \gamma = \frac{1-c}{c} \int^{ \gamma_c}_{ 0} \tau   d \gamma=  \frac{1-c}{c}\psi_1^e .
\label{2-184}
\eey
Substituting Eq. (\ref{2-184}) in Eq. (\ref{2-182}), we obtain
\bey
X=   \psi_1^e  - \frac{\Delta \psi^\theta  h_b+ 2  \Gamma_0}{h}\geq 0,
\label{2-185}
\eey
which is equivalent to Eq. (\ref{2-180}).

\end{document}